\documentclass[11pt,a4paper]{article}

\usepackage[utf8]{inputenc}
\usepackage[T1]{fontenc}
\usepackage[english]{babel}
\usepackage{lmodern}
\usepackage{microtype}
\usepackage{geometry}
\usepackage{graphicx}
\usepackage{booktabs}
\usepackage{tabularx}
\usepackage{hyperref}
\usepackage{enumitem}
\usepackage{tikz}
\usetikzlibrary{arrows.meta,positioning,shapes.geometric}

\hypersetup{
  colorlinks=true,
  linkcolor=blue,
  citecolor=blue,
  urlcolor=blue
}

\newcommand{\doi}[1]{\href{https://doi.org/#1}{doi:\ \nolinkurl{#1}}}

\title{Reversa: A Reverse Documentation Engineering Framework for Converting Legacy Software into Operational Specifications for AI Agents}
\author{Sanderson Oliveira de Macedo\\
\textit{Federal Institute of Goias - Brazil}\\
\texttt{sanderson.macedo@ifg.edu.br}\\
\href{https://orcid.org/0000-0002-5255-596X}{ORCID: 0000-0002-5255-596X}
\and
Ronaldo Martins da Costa\\
\textit{Federal University of Goias - Brazil}\\
\texttt{ronaldocosta@ufg.br}\\
\href{https://orcid.org/0000-0003-1892-9080}{ORCID: 0000-0003-1892-9080}}
\date{May 18, 2026}

\begin{document}

\maketitle

\begin{abstract}
Legacy systems concentrate business rules, architectural decisions, and operational exceptions that often remain implicit in code, data, configuration, and maintenance practices. At the same time, language-model-based coding agents depend on reliable context, correctness criteria, and behavioral contracts to modify real systems with lower risk. This paper presents Reversa, a reverse documentation engineering framework for converting legacy software into traceable operational specifications for AI agents. Reversa organizes this process as a multi-agent pipeline: specialized agents map the project surface, analyze modules, extract implicit rules, synthesize architecture, write unit-level specifications, and review generated claims. The proposal emphasizes three mechanisms: traceability between code and specification, explicit confidence marking, and preservation of gaps for human validation. The framework is distributed as a Node.js CLI, installs skills across multiple agent engines, and uses a SHA-256 manifest to preserve modified files during update or uninstall operations. In addition to the architectural description, we report an exploratory case study on migrating an ATM from COBOL to Go, in which the pipeline produced 517 claims classified by an internal confidence index, 10 registered gaps, 53 Gherkin parity scenarios, and a reconstruction plan with 9 of 11 tasks completed at inventory time. Final parity validation and cutover were not completed in this study. We do not claim broad empirical superiority; we position the contribution with respect to the literature on reverse engineering, LLM-based documentation, and software agents, and propose an evaluation protocol with metrics for coverage, traceability, confidence, utility, and cost.
\end{abstract}


\section{Introduction}
\label{sec:introduction}

Legacy systems are rarely only old collections of source files. In many contexts, they concentrate business rules, architectural decisions, operational exceptions, data conventions, and usage flows accumulated over years of maintenance, modernization, and evolution \cite{washizaki2018prometa,reis2017Xis,assuncao2024modernization,mohottige2025microservices,diggs2024legacy,luo2024repoagent}. Part of this knowledge appears in function names, SQL queries, validations, comments, or configuration files; another part remains implicit in implementation patterns, change history, actual system use, or tacit knowledge. Therefore, modifying, modernizing, replatforming, or migrating the programming language of a legacy system does not only require editing code; it requires reconstructing, with some degree of confidence, what the system does, why it does it, and which behaviors must not be broken.

Without a systematic bridge between legacy systems without explicit specifications and agents that need operational contracts, AI-assisted maintenance automation tends to operate on incomplete context and implicit confidence. Recent literature shows that intermediate representations, repository-level documentation, and specification generation remain central to understanding, comparing, and maintaining software \cite{washizaki2018prometa,luo2024repoagent,athale2025Knowledge,xie2025specifications,krishna2024requirements,hou2024llmse}. At the same time, AI agents already execute real tasks in repositories and require their own interfaces to operate effectively \cite{jimenez2024swebench,yang2024sweagent,li2026Reproducible,liu2026agentsurvey}. Despite these advances, it remains open how to transform implicit legacy knowledge into traceable operational specifications, with explicit uncertainty, to guide coding agents.

In this work, we propose Reversa, a reverse documentation engineering framework aimed at legacy systems and coding agents. We call \textit{reverse documentation engineering} the process of deriving, from an existing system, technical documentation and operational specifications that make behavior, architecture, domain rules, gaps, and confidence levels explicit. Reversa organizes this process as a multi-agent pipeline: a central orchestrator coordinates specialized agents to map the project surface, analyze modules, extract implicit rules, synthesize architecture, write unit-level specifications, and review generated claims. In addition to describing the framework, we present an exploratory case study in which Reversa supports the migration of an ATM from COBOL to a Go reimplementation.

The expected result is not only narrative documentation, but operational specifications that are traceable to code, marked by confidence levels, and accompanied by gaps that require human validation. The terminology is deliberate: ``reverse engineering'' alone often evokes recovery of code or architecture from binaries; here, the goal is to transform an existing system into documentary contracts that guide AI-assisted maintenance, migration, and evolution. This emphasis also delimits the scope of the paper: Reversa is not presented as a replacement for human validation, nor as definitive empirical evidence of superiority over other approaches, but as a system framework, an evaluation protocol, and a first exploratory evidence point.

This paper makes five contributions: (i) we operationally define reverse documentation engineering in the context of legacy systems and AI agents; (ii) we present Reversa as a multi-agent framework for converting legacy code into traceable operational specifications; (iii) we describe a confidence and gaps model for reducing the risk of generated documentation with false certainty; (iv) we detail an architecture portable across coding-agent engines, with installation and update controlled by a manifest; and (v) we propose an evaluation protocol and instantiate it in an exploratory COBOL--Go case study.

These contributions are guided by four research questions:

\begin{description}[leftmargin=1.2cm,style=nextline]
  \item[RQ1] How can a multi-agent framework transform legacy code into operational specifications consumable by AI agents?
  \item[RQ2] Which traceability and confidence mechanisms are needed to prevent agent-generated specifications from hiding uncertainty about legacy behavior?
  \item[RQ3] How does the separation of agentic roles help cover distinct stages of understanding, synthesis, specification writing, and review?
  \item[RQ4] What evaluation protocol can measure whether generated specifications are useful for AI-assisted maintenance, migration, or evolution?
\end{description}

The remainder of the paper is organized as follows. Section~\ref{sec:related-work} discusses related work; Section~\ref{sec:reversa-framework} presents the framework; Section~\ref{sec:evaluation-methodology} describes the evaluation methodology; Section~\ref{sec:atm-case-study} reports the exploratory case study; Section~\ref{sec:discussion} discusses implications; Section~\ref{sec:threats-validity} presents threats to validity; and Section~\ref{sec:conclusion} concludes the paper.

\section{Background and Related Work}
\label{sec:related-work}

This section positions Reversa across four research lines: reverse engineering and program comprehension; repository documentation and summarization with LLMs; requirements and specification generation; and LLM-based software agents. The synthesis of these lines shows that the central problem is not the absence of techniques for understanding software, generating documentation, or executing tasks with agents, but the lack of a systematic bridge between legacy systems without explicit specifications and traceable operational specifications for coding agents.

\subsection{Reverse engineering and program comprehension}

Reversa builds on recent reverse-engineering, architecture-description, and modernization work that treats existing systems as evidence sources for reconstructing structure, behavior, and modernization options. Program metamodels and intermediate representations remain important for classifying and extending reverse-engineering tools \cite{washizaki2018prometa}. Model-driven reverse engineering for legacy information systems and contemporary modernization studies reinforce the need to preserve domain knowledge while changing technology platforms \cite{reis2017Xis,assuncao2024modernization,mohottige2025microservices}. This framing delimits the proposal of this paper: Reversa does not automatically recompile, restructure, or modernize the system; it derives technical documentation and operational specifications from existing evidence.

Recent empirical work on architecture practice shows that software architecture activities remain difficult across requirements, design, construction, testing, and maintenance, with recurring challenges around knowledge management, documentation, tooling, and process \cite{wan2023architecture}. This view is compatible with the Reversa hypothesis: before delegating maintenance, migration, or evolution to AI agents, operational knowledge about the system must be reconstructed. Reversa aligns with this modern reverse-engineering and modernization tradition, but changes the primary consumer of the artifact: from humans and analysis tools to coding agents that need actionable, traceable, and reviewable contracts.

\subsection{Repository documentation and summarization with LLMs}

Recent work shows that LLMs can support documentation and comprehension at repository level. RepoAgent proposes a framework to generate, maintain, and update repository-level code documentation \cite{luo2024repoagent}. Hierarchical summarization approaches also explore summaries at multiple levels, aggregating information from functions, files, and packages to support comprehension of business applications \cite{dhulshette2025Hierarchical}. Work on LLM-generated documentation for legacy modernization explicitly studies legacy languages and highlights both the promise of generated comments and the need for better evaluation metrics \cite{diggs2024legacy}. Knowledge-graph and benchmark-oriented work further indicates that repository-level context is becoming a first-class object for code generation and evaluation \cite{athale2025Knowledge,zhang2025Repocbench}. These works are close to Reversa because they treat the repository as the unit of analysis, not only isolated functions.

However, generated documentation is not automatically safe as a basis for agents. Macke and Doyle empirically show that incorrect documentation can harm LLM code understanding, while missing or incomplete documentation does not necessarily produce the same type of harm \cite{macke2024Testing}. This result supports a central Reversa decision: documentary claims should carry explicit confidence levels and gaps, rather than presenting fragile inferences as facts. Thus, the goal of Reversa is not only to produce explanatory text, but to organize reverse documentation as an operational specification that is traceable to code and suitable for human validation.

\subsection{Requirements and specification generation with LLMs}

Another line close to Reversa investigates requirements and specification generation with LLMs. Xie et al. evaluate LLMs for generating software specifications from comments or documentation, comparing multiple models with traditional approaches and analyzing failure cases \cite{xie2025specifications}. Krishna et al. evaluate GPT-4 and CodeLlama for generating, validating, and correcting software requirements specification documents \cite{krishna2024requirements}. ReqInOne, in turn, proposes a modular agent for converting natural language into structured SRS artifacts through stages such as summarization, extraction, and requirements classification \cite{zhu2025reqinone}. Adjacent requirements-engineering work explores generative AI, retrieval-augmented generation, use-case model generation, and cause-effect graph generation as mechanisms to improve requirements and test-related artifacts \cite{arora2024Advancing,masoudifard2024Integrating,eisenreich2025Leveraging,kirinuki2024Cegen}.

These works demonstrate that LLMs can produce useful specification artifacts, but they differ from Reversa in two aspects. First, many start from requirements, comments, or documentation that are already available, while Reversa starts from existing systems in which operational knowledge may be only implicit in code, data, and implementation conventions. Second, specification in Reversa is oriented toward use by coding agents, with traceability, confidence, and explicit gaps as part of the contract. Formal-specification and assertion-generation works such as AutoReSpec, KerSpecGen, and RAG-driven assertion generation reinforce the relevance of verifiable specifications and validation feedback \cite{ayon2026autorespec,wang2025kerspecgen,liu2025Rag}, but have a more localized or specialized scope than reverse documentation engineering for whole legacy systems.

\subsection{Software agents and multi-agent workflows}

LLM-based agents are already an active front in software engineering. Recent surveys map hundreds of works on LLMs for SE and LLM-based agents, covering tasks such as code generation, testing, debugging, maintenance, tool use, and interaction with external environments \cite{zheng2023codesurvey,gormez2024mapping,hou2024llmse,terragni2024Future,liu2026agentsurvey}. Work on code models, program synthesis, code-generation challenges, reasoning and planning, agent training, and practical prompting further shows that the agentic SE landscape is broadening beyond single-turn code completion \cite{chen2021codex,austin2021program,wang2024Advancements,ding2024Reasoning,wang2024Learning,wienholt2025Prompt}. SWE-bench shows that real maintenance problems extracted from issues and pull requests require repository comprehension, coordination across files, and validation through tests \cite{jimenez2024swebench}. SWE-agent, in turn, demonstrates that agent-computer interfaces influence the performance of agents that navigate, edit, and test repositories \cite{yang2024sweagent}. Recent evaluation work also stresses that agentic AI for software engineering needs reproducible, explainable, and effective evaluation settings \cite{li2026Reproducible}; broader accounts of LLMs in software engineering and acquisition reinforce the need to connect automation with governance, evidence, and process constraints \cite{zhao2023llms,robert2026Transforming,kessel2024Morescient}.

These works establish the downstream consumer of Reversa: agents that act on real code. However, they usually assume that the task, context, and correctness criteria are already available. Reversa proposes an earlier or complementary layer: transforming implicit legacy knowledge into operational specifications that can guide those agents before maintenance, migration, or evolution is executed. Multi-agent works such as ChatDev show that role decomposition and communication between agents can structure development workflows \cite{qian2024chatdev}; in Reversa, this decomposition is applied to another life-cycle phase, involving system mapping, technical analysis, rule extraction, architectural synthesis, specification writing, and confidence review.

\subsection{Positioning synthesis}

The literature covers important parts of the problem: reverse engineering provides the basis for recovering knowledge from existing systems; documentation and summarization with LLMs show how to produce repository-level explanations; requirements and specification generation explores formal or semi-formal artifacts; and software agents demonstrate execution of real tasks in repositories. Even so, the intersection of these fronts remains underconsolidated: converting legacy software without explicit specifications into traceable operational contracts, marked by confidence and gaps, to guide coding agents.

\begin{table}[t]
  \centering
  \footnotesize
  \begin{tabular}{p{2.9cm} p{3.4cm} p{3.2cm} p{3.2cm}}
    \toprule
    \textbf{Work line} & \textbf{Main focus} & \textbf{Typical consumer} & \textbf{Reversa position} \\
    \midrule
    Classical reverse engineering & Recover components, relationships, abstractions, and design. & Engineers, architects, and analysis tools. & Uses the conceptual basis, but directs the output to coding agents. \\
    Repository documentation with LLMs & Generate and maintain explanations at code and repository level. & Human developers and technical documentation. & Requires traceability, confidence, and gaps for operational use. \\
    Requirements and specification generation & Produce SRS, formal contracts, or semi-formal specifications. & Analysts, verifiers, and requirements pipelines. & Uses the legacy system as the primary source, not only requirements or comments. \\
    Software agents & Execute maintenance, testing, debugging, and development tasks. & Repositories with defined tasks and correctness criteria. & Provides an earlier layer of operational contracts for these agents. \\
    \bottomrule
  \end{tabular}
  \caption{Comparison between related-work lines and the positioning of Reversa.}
  \label{tab:related-work-comparison}
\end{table}

Reversa occupies this intersection. Its contribution is not to replace classical reverse engineering, automatic documentation, SRS generation, or coding agents, but to connect them through a reverse documentation engineering pipeline. This pipeline treats legacy code as a source of evidence, produces operational specifications as an intermediate artifact, and preserves human validation as an explicit part of the process.

\section{The Reversa Framework}
\label{sec:reversa-framework}

This section presents Reversa as a reverse documentation engineering framework for legacy systems. The objective of the framework is not to replace static analyzers, automatic documentation tools, or existing coding agents. Its function is to produce an intermediate layer of operational specifications: artifacts traceable to code, marked by confidence levels, and organized to guide agents that will maintain, migrate, or evolve the system. The description below is based on the project snapshot analyzed in this paper, in which Reversa is distributed as a Node.js package with a CLI, installable agents, and artifact templates.

\subsection{Design goals}
\label{subsec:design-goals}

The design of Reversa starts from five goals. First, the framework must produce operational contracts, not only narrative explanations. An operational specification should record expected behavior, domain rules, dependencies, flows, reimplementation tasks, and sufficient evidence for another agent or developer to act on the system with less ambiguity. Second, every relevant claim should be traceable to legacy evidence, such as files, modules, routes, schemas, queries, or previously extracted artifacts.

Third, the framework must make uncertainty explicit. Instead of hiding fragile inferences in declarative text, Reversa separates confirmed claims, inferred claims, and gaps that require human validation. Fourth, installation must not destroy or take over the legacy project: installed files are registered, versioned, and protected by a manifest. Fifth, the process must be portable across coding-agent engines, because the problem addressed by Reversa precedes the choice of a specific tool.

\subsection{General architecture}
\label{subsec:general-architecture}

Reversa is implemented as a JavaScript CLI, executed through the \texttt{npx reversa} command. The entry point \texttt{bin/reversa.js} dispatches commands such as \texttt{install}, \texttt{update}, \texttt{status}, \texttt{uninstall}, \texttt{add-agent}, \texttt{add-engine}, and \texttt{export-diagrams}. The standard installation starts with \texttt{npx reversa install}. This command detects engines available in the project, asks which agent teams should be installed, collects project metadata, and writes the state structure under \texttt{.reversa/}.

The architecture separates three layers. The first is the installation and preservation layer, composed of engine detection, prompts, file writing, validation, and manifest handling. The second is the agent layer, formed by installable skills with their own instructions, references, and templates. The third is the artifact layer, normally materialized in \texttt{\_reversa\_sdd/}, where inventories, analyses, unit-level specifications, traceability matrices, gaps, and confidence reports are written. This artifact layer follows contemporary architecture-description practice: system knowledge should be expressed through structured descriptions, concerns, viewpoints, and complementary artifacts rather than through a single diagram or prose document \cite{iso42010,bass2021architecture,wan2023architecture}.

\begin{figure}[t]
  \centering
  \includegraphics[width=\linewidth]{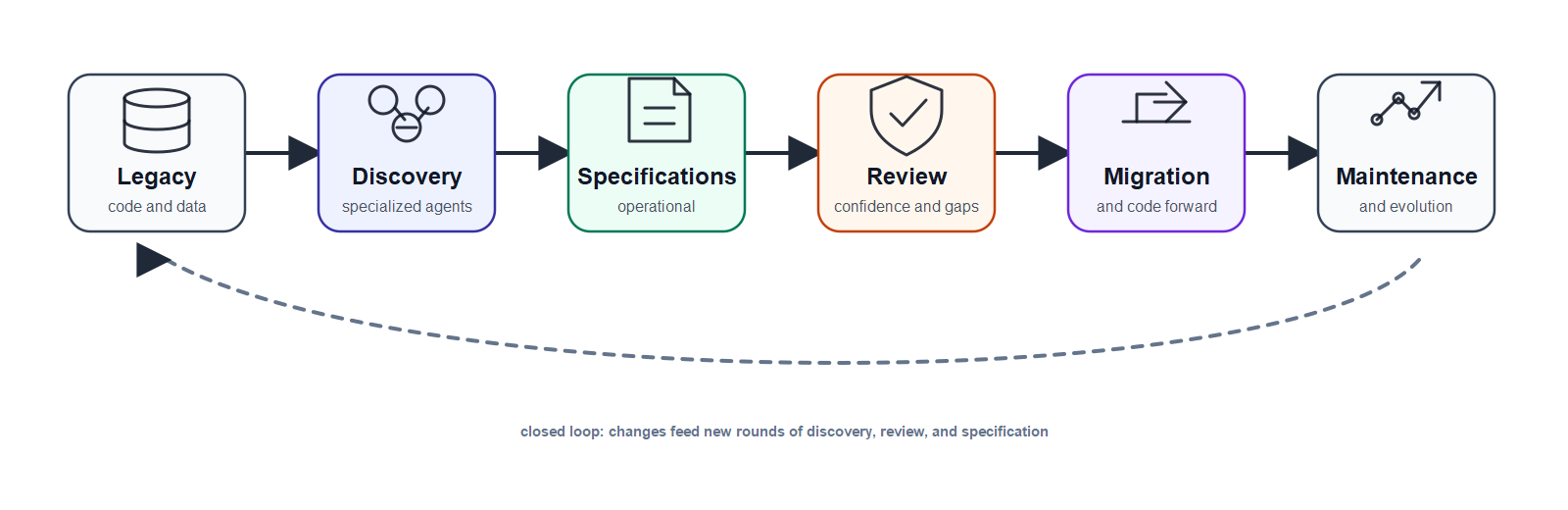}
  \caption{Conceptual pipeline of Reversa. The legacy system is analyzed by specialized agents; the resulting specifications are reviewed for confidence and gaps; and the artifacts then guide future migration, maintenance, and evolution.}
  \label{fig:reversa-pipeline}
\end{figure}

Figure~\ref{fig:reversa-pipeline} summarizes the flow. The legacy system is the primary source of evidence; the Discovery team transforms this evidence into artifacts; the review process reclassifies fragile claims and makes questions to the user explicit; and downstream teams consume the specifications for migration or future development. This structure addresses RQ1 by showing how the transformation of legacy code into an operational specification is decomposed into steps executable by agents.

\subsection{Portable installation and manifest-based preservation}
\label{subsec:installation-manifest}

During installation, Reversa detects engines such as Claude Code, Codex, Cursor, Gemini CLI, Windsurf, Antigravity, Kiro, Opencode, Cline, Roo Code, GitHub Copilot, Aider, and Amazon Q Developer. For each selected engine, the installer copies the agent skills to the expected directory and installs an appropriate entry file, such as \texttt{AGENTS.md}, \texttt{CLAUDE.md}, \texttt{GEMINI.md}, or engine-specific rules. This decision makes the framework independent of a single agent interface.

The \texttt{.reversa/} directory concentrates the operational state. It includes \texttt{state.json}, \texttt{config.toml}, \texttt{config.user.toml}, \texttt{plan.md}, a version file, templates, scripts, and installation metadata. In addition, Reversa generates a SHA-256 manifest at \texttt{.reversa/\_config/files-manifest.json}. This manifest records hashes for installed files and enables each file to be classified as intact, modified, or missing. During update, intact or missing files can be updated, while files modified by the user are preserved. During uninstall, the same principle reduces the risk of deleting content that does not belong to Reversa or was locally changed.

\subsection{Agent teams}
\label{subsec:agent-teams}

The main abstraction in Reversa is a team of agents with separated responsibilities. The Discovery team covers the path from initial mapping to specification writing and review. Additional teams extend the cycle to migration, future evolution, pricing, and translation of structured inputs. Table~\ref{tab:reversa-agents} summarizes the central roles used in this paper.

\begin{table}[t]
  \centering
  \footnotesize
  \begin{tabularx}{\linewidth}{p{3.0cm} X X}
    \toprule
    \textbf{Agent or team} & \textbf{Main role} & \textbf{Expected artifacts} \\
    \midrule
    \texttt{reversa} & Orchestrates execution, resumption, and activation of other agents. & Execution plan and state under \texttt{.reversa/}. \\
    \texttt{reversa-scout} & Maps project surface, stack, dependencies, and entry points. & Initial inventory and evidence-backed signals. \\
    \texttt{reversa-}\newline\texttt{archaeologist} & Deepens technical analysis of modules and internal structures. & Modular analysis and classified technical facts. \\
    \texttt{reversa-detective} & Extracts business rules, states, permissions, and implicit exceptions. & Recovered rules, gaps, and decisions. \\
    \texttt{reversa-architect} & Synthesizes architecture, dependencies, data, and impact across components. & Architecture, diagrams, and impact matrix. \\
    \texttt{reversa-writer} & Converts extracted knowledge into unit-level specifications. & \texttt{requirements.md}, \texttt{design.md}, \texttt{tasks.md}, and traceability. \\
    \texttt{reversa-reviewer} & Reviews claims, confidence, and gaps for human validation. & Confidence report, questions, and remaining gaps. \\
    Migration & Plans reconstruction or modernization from the specifications. & Strategy, risks, target architecture, and parity specs. \\
    Code Forward & Transforms new ideas into requirements, roadmap, actions, and audit. & Evolution artifacts traced to the legacy system. \\
    \bottomrule
  \end{tabularx}
  \caption{Central agentic roles in Reversa and their expected artifacts.}
  \label{tab:reversa-agents}
\end{table}

This decomposition addresses RQ3. Instead of asking a single agent to read the whole repository and produce final documentation in one step, Reversa distributes the process across roles with explicit inputs and outputs. This separation reduces cognitive coupling between tasks: mapping the project surface differs from inferring domain rules; synthesizing architecture differs from writing unit-level contracts; reviewing confidence differs from generating new text.

\subsection{Confidence, gaps, and traceability model}
\label{subsec:confidence-gaps}

The confidence model is the main uncertainty-management mechanism in Reversa. Agent instructions distinguish confirmed claims, inferred claims, and gaps. A confirmed claim must have direct evidence in code or in a verifiable artifact. An inferred claim may be supported by recurring patterns, names, flows, or structure, but should not be written as certainty. A gap records information that could not be determined safely and needs human validation.

This model appears especially in the Writer and Reviewer. The Writer must produce specifications with evidence and traceability, omitting lines when there is not enough basis. The Reviewer rereads the specifications, returns to the original code to check fragile claims, reclassifies items when necessary and consolidates user questions in artifacts such as \texttt{questions.md}, \texttt{gaps.md}, and \texttt{confidence-report.md}. At more complete documentation levels, the process also generates matrices such as \texttt{traceability/code-spec-matrix.md} and \texttt{traceability/spec-impact-matrix.md}. This emphasis is consistent with recent mapping evidence that traceability can affect maintenance and evolution activities, while also introducing maintenance costs that must be managed \cite{tian2021traceability}.

This design addresses RQ2. The point is not to claim that agents never err, but to create a protocol in which uncertainty is preserved as output data. For legacy systems, this decision is central: a partially uncertain specification that is honest about its gaps is more useful for responsible maintenance than fluent documentation that presents assumptions as facts.

\subsection{Closed loop: discovery, migration, and evolution}
\label{subsec:closed-loop}

Reversa also proposes a closed loop between discovery, migration, and evolution. The first cycle extracts knowledge from the legacy system and produces operational specifications. The Migration team consumes these artifacts to plan modernization, record paradigm decisions, select strategy, map risks, design the target architecture, model domain and data, and produce parity specifications. The Code Forward team acts in another direction: it starts from a new feature intent, transforms this intent into requirements and a plan, audits consistency with existing specifications, and guides future implementation.

\begin{figure}[t]
  \centering
  \includegraphics[width=\linewidth]{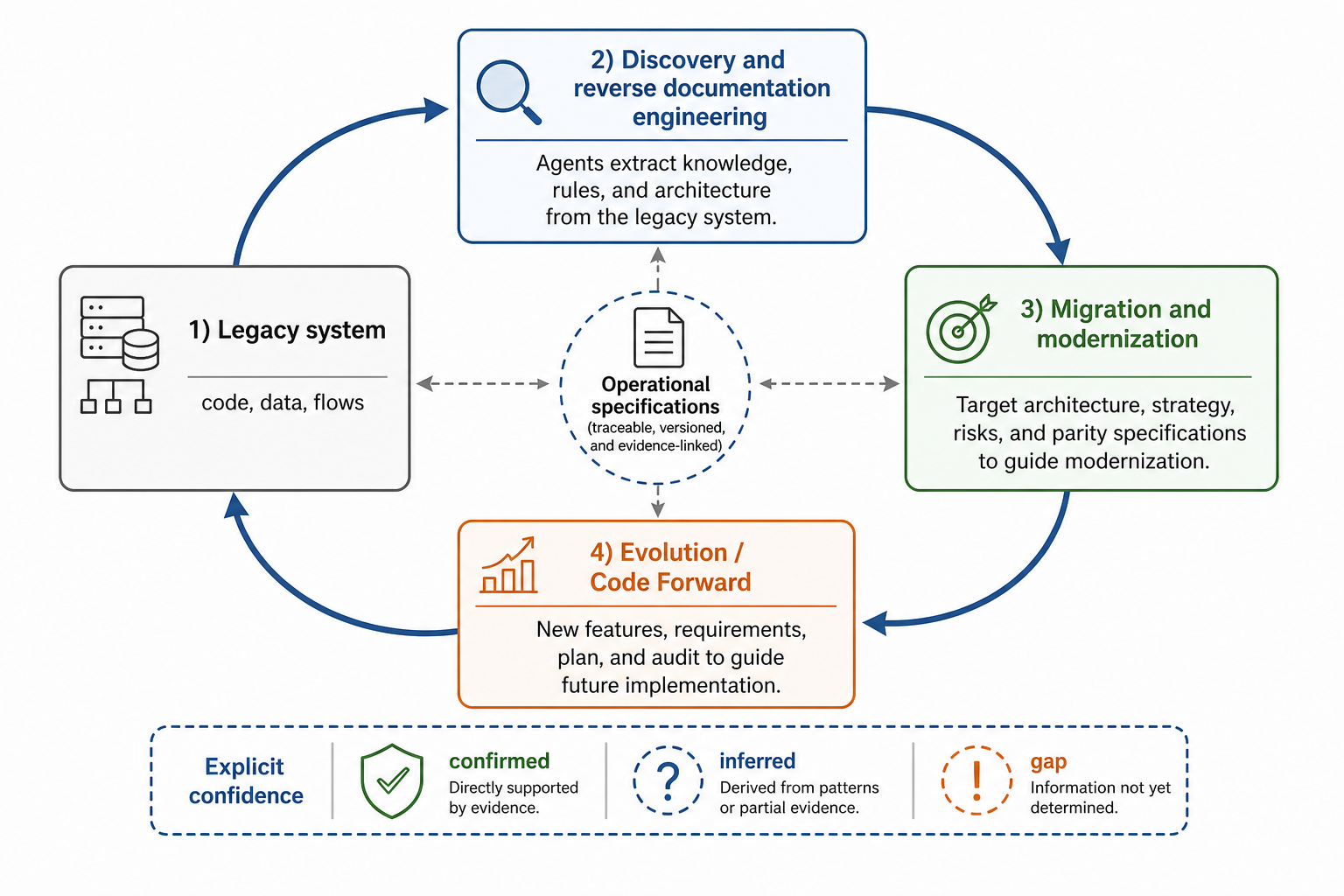}
  \caption{Closed loop in Reversa: discovery recovers legacy knowledge, migration transforms it into modernization decisions, and Code Forward reuses it as a basis for continuous evolution. Explicit confidence, with confirmed, inferred, and gap states, crosses the whole flow.}
  \label{fig:closed-loop}
\end{figure}

This connection differentiates Reversa from a tool that only generates static documentation. Reverse documentation becomes a maintained operational artifact: it guides changes and, after those changes, can be used as a reference for new extractions, regression checks, and contract updates. In this version of the paper, this hypothesis is presented as a design contribution and initial exploratory evidence, not as a broad empirical conclusion. Section~\ref{sec:evaluation-methodology} defines how this cycle can be evaluated, and Section~\ref{sec:atm-case-study} reports a first instantiation of the protocol.

\section{Evaluation Methodology}
\label{sec:evaluation-methodology}

This section defines how to evaluate the extent to which operational specifications generated by reverse documentation engineering are traceable, reliable, and potentially useful to support AI-assisted maintenance, migration, and evolution. The protocol is informed by prior evaluations of specification generation, repository-level code tasks, and agentic software-engineering systems \cite{xie2025specifications,krishna2024requirements,jimenez2024swebench,yang2024sweagent,li2026Reproducible}. In this version, the protocol is instantiated in an exploratory case study, reported in Section~\ref{sec:atm-case-study}. The evaluation observes initial plausibility and traceability; controlled comparisons with baselines and direct measurements of utility remain future work.

The protocol has five stages. First, record the initial state of the project: language, dependencies, existing documentation, presence of tests, and analyzed version. Second, install Reversa with \texttt{npx reversa install}, selecting engines, agent teams, and output configuration. Third, execute the Discovery team until the main artifacts are generated in \texttt{\_reversa\_sdd/}. Fourth, review confidence and gaps, collecting \texttt{confidence-report.md}, \texttt{questions.md}, \texttt{gaps.md}, and traceability matrices when available. Fifth, evaluate the artifacts through human inspection and, when possible, through downstream tasks executed by coding agents.

The main unit of analysis is the legacy project submitted to Reversa. Within each project, the evaluation may observe modules, endpoints, screens, entities, business flows, or services, according to the chosen granularity. In the exploratory study in this version, the unit of analysis is an ATM in COBOL and the internal units are the modules used in the parity scope.

\begin{table}[t]
  \centering
  \footnotesize
  \begin{tabular}{p{2.7cm} p{4.4cm} p{3.1cm} p{2.4cm}}
    \toprule
    \textbf{Metric} & \textbf{Description} & \textbf{Expected interpretation} & \textbf{In the ATM case} \\
    \midrule
    File coverage & Proportion of relevant files associated with some specification. & Measures extraction reach, but does not guarantee quality without audit. & Observed by inventory. \\
    Unit coverage & Proportion of modules, endpoints, screens, or entities with their own specifications. & Indicates alignment between system organization and chosen granularity. & Observed by module. \\
    Traceability density & Average number of evidence references per requirement, rule, or task. & Suggests auditability when the evidence is pertinent. & Observed qualitatively. \\
    Confidence distribution & Proportion of confirmed claims, inferred claims, and gaps. & Exposes internal certainty classification; does not measure factual accuracy without audit. & Measured. \\
    Blocking gaps & Number of gaps that prevent safe reimplementation, migration, or maintenance. & Helps prioritize human validation before downstream tasks. & Measured. \\
    Expert precision & Percentage of claims accepted by independent human reviewers. & Measures substantive quality of the specifications. & Not measured. \\
    Agent utility & Performance difference between agents with and without Reversa specifications. & Evaluates support for AI-assisted maintenance, migration, or evolution. & Not measured in a controlled way. \\
    \bottomrule
  \end{tabular}
  \caption{Candidate metrics for evaluating operational specifications generated by Reversa and observation status in the ATM study.}
  \label{tab:evaluation-metrics}
\end{table}

In addition to these metrics, operational cost should be recorded: execution time, number of human interactions, quantity of generated questions, artifact size, traceability maintenance effort, and review effort. A specification that is useful but too expensive to maintain may not be viable in real teams, especially because traceability itself introduces both benefits and maintenance costs \cite{tian2021traceability}.

To make the evaluation auditable, each case study should produce a minimal artifact package: project identifier, analyzed version, Reversa configuration, engines and teams used, main artifacts under \texttt{\_reversa\_sdd/}, confidence and gap reports, relevant instructions, parity scenarios when they exist, and a downstream task plan. When human intervention occurs, the package should also record decisions made, answered questions, and gaps removed from scope.

The most relevant baselines are: (i) a single repository documentation prompt, without role decomposition; (ii) a generic coding agent operating without prior specifications; and (iii) a conventional documentation or static-analysis tool, when applicable. Repository-documentation, requirements-generation, and coding-agent benchmarks provide useful reference points for designing these comparisons \cite{luo2024repoagent,zhang2025Repocbench,zhu2025reqinone,jimenez2024swebench}. Reversa should not be compared by text volume, but by utility, traceability, explicit uncertainty, and capacity to guide subsequent actions.

In the ATM case, we recorded the transformation of artifacts into a reconstruction plan and parity tests, but we have not yet executed a controlled comparison with a single agent or an independent factual-precision review. Thus, the initial study observes part of the metrics in a realistic scenario, while confirmation of the hypothesis requires new projects, independent review, and comparison with baselines.

\section{Exploratory Case Study: ATM COBOL to Go}
\label{sec:atm-case-study}

To instantiate the evaluation protocol in a concrete scenario, we analyze an exploratory case study on migrating an ATM system written in COBOL to a Go reimplementation. The objective of the study is not to demonstrate statistical generalization, but to observe whether Reversa can produce traceable artifacts, explicit gaps, and operational inputs capable of structuring a controlled reconstruction. The case was conducted in an educational project in a simplified banking domain, with a single stakeholder and no production use.

\subsection{Legacy characterization}
\label{subsec:atm-characterization}

The legacy system, called \texttt{banco-atm}, implements basic automated teller machine operations in GnuCOBOL. The active parity scope includes five modules: \texttt{MENU}, \texttt{CONTA}, \texttt{EXTRATO}, \texttt{UTIL}, and \texttt{kbdread}. Two components were excluded from parity: \texttt{ADD-CLIENTE}, because it is a technical operation outside the ATM domain, and \texttt{x25-communication}, because it exists as a specification without a corresponding implementation. The original persistence uses \texttt{.DAT} files; no automated tests, CI/CD, or Docker setup existed in the legacy system.

\begin{table}[t]
  \centering
  \footnotesize
  \begin{tabular}{p{4.0cm} p{8.4cm}}
    \toprule
    \textbf{Aspect} & \textbf{Observed value} \\
    \midrule
    Domain & Single-user, single-process ATM for basic checking-account operations. \\
    Languages & COBOL, with a C helper for unbuffered keyboard reading. \\
    Modules in scope & \texttt{MENU}, \texttt{CONTA}, \texttt{EXTRATO}, \texttt{UTIL}, and \texttt{kbdread}. \\
    Persistence & Indexed and sequential \texttt{.DAT} files. \\
    Preexisting tests & No automated tests found. \\
    Target system & Go reimplementation with SQLite and Gherkin parity tests. \\
    \bottomrule
  \end{tabular}
  \caption{Summary characterization of the COBOL-to-Go ATM case study.}
  \label{tab:atm-characterization}
\end{table}

\subsection{Pipeline execution}
\label{subsec:atm-pipeline-execution}

The Reversa pipeline was executed between May 4 and May 7, 2026. The discovery stage produced artifacts such as inventory, code analysis, architecture, domain model, state machines, dependencies, questions, gaps, and confidence report. The migration stage produced a briefing, paradigm and topology decisions, target business rules, migration strategy, risk register, cutover plan, target architecture, domain model, data model, data migration plan, parity specifications, Gherkin tests, and coding handoff.

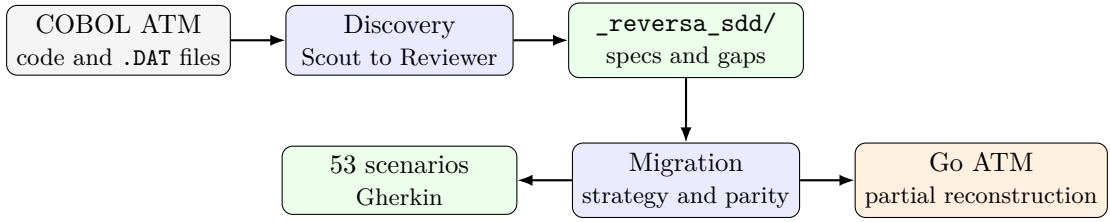
\begin{figure}[t]
  \centering
  \begin{tikzpicture}[
    font=\small,
    node distance=0.72cm,
    box/.style={draw, rounded corners, align=center, minimum width=2.8cm, minimum height=0.9cm, fill=gray!8},
    agent/.style={draw, rounded corners, align=center, minimum width=3.0cm, minimum height=0.9cm, fill=blue!8},
    artifact/.style={draw, rounded corners, align=center, minimum width=3.1cm, minimum height=0.9cm, fill=green!8},
    gate/.style={draw, rounded corners, align=center, minimum width=2.8cm, minimum height=0.9cm, fill=orange!12},
    arrow/.style={-{Latex[length=2mm]}, thick}
  ]
    \node[box] (legacy) {COBOL ATM\\\footnotesize code and \texttt{.DAT} files};
    \node[agent, right=of legacy] (discovery) {Discovery\\\footnotesize Scout to Reviewer};
    \node[artifact, right=of discovery] (sdd) {\texttt{\_reversa\_sdd/}\\\footnotesize specs and gaps};
    \node[agent, below=0.85cm of sdd] (migration) {Migration\\\footnotesize strategy and parity};
    \node[artifact, left=of migration] (tests) {53 scenarios\\\footnotesize Gherkin};
    \node[gate, right=of migration] (go) {Go ATM\\\footnotesize partial reconstruction};

    \draw[arrow] (legacy) -- (discovery);
    \draw[arrow] (discovery) -- (sdd);
    \draw[arrow] (sdd) -- (migration);
    \draw[arrow] (migration) -- (tests);
    \draw[arrow] (migration) -- (go);
  \end{tikzpicture}
  \caption{Instantiation of the Reversa pipeline in the ATM case study. The COBOL legacy system feeds discovery, the artifacts in \texttt{\_reversa\_sdd/} guide migration, and the parity specifications guide the Go reconstruction.}
  \label{fig:atm-case-study}
\end{figure}

Figure~\ref{fig:atm-case-study} summarizes the execution. The most important intermediate result was the transition from a system without tests to a set of specification artifacts and 53 Gherkin parity scenarios, distributed across login, balance, withdrawal, deposit, transfer, statement, monetary formatting, and keyboard masking. These scenarios act as a bridge between reverse documentation and executable reconstruction.

\subsection{Confidence, gaps, and reconstruction}
\label{subsec:atm-confidence}

The confidence report classified 517 claims in the active scope. Of this total, 490 were marked as confirmed, 24 as inferred, and 3 as gaps. Using the operational rule adopted in the study, where confirmed claims are worth 1.0 and inferred claims are worth 0.5, the calculated internal confidence index was 97.1\%. This index summarizes the classification assigned by the pipeline and should not be interpreted as factual precision, because there was no external audit of the claims. Table~\ref{tab:atm-confidence} presents the distribution by module.

\begin{table}[t]
  \centering
  \footnotesize
  \begin{tabular}{lrrrr}
    \toprule
    \textbf{Spec} & \textbf{Confirmed} & \textbf{Inferred} & \textbf{Gap} & \textbf{Index} \\
    \midrule
    \texttt{menu} & 32 & 1 & 0 & 98.5\% \\
    \texttt{conta} & 129 & 6 & 0 & 97.8\% \\
    \texttt{extrato} & 124 & 9 & 1 & 95.9\% \\
    \texttt{util} & 115 & 4 & 2 & 96.7\% \\
    \texttt{kbdread} & 90 & 4 & 0 & 97.9\% \\
    \midrule
    Active total & 490 & 24 & 3 & 97.1\% \\
    \bottomrule
  \end{tabular}
  \caption{Distribution of the internal confidence index for claims in the active scope of the case study.}
  \label{tab:atm-confidence}
\end{table}

In addition to claim-level gaps, the process registered 10 project gaps: 3 critical, 3 moderate, 2 cosmetic, and 2 out of scope. Five critical or moderate gaps were resolved by documented human decision; three remained residual; and two were removed from the parity scope. This result illustrates the expected function of Reversa: not only to increase the amount of documentation, but to transform operational uncertainty into visible, prioritizable, and traceable items.

\begin{table}[t]
  \centering
  \footnotesize
  \begin{tabular}{lrl}
    \toprule
    \textbf{Gap class} & \textbf{Quantity} & \textbf{Treatment in the study} \\
    \midrule
    Critical & 3 & Resolved by documented human decision. \\
    Moderate & 3 & Two resolved; one kept as residual. \\
    Cosmetic & 2 & Kept as low-severity residuals. \\
    Out of scope & 2 & Excluded from the parity scope. \\
    \midrule
    Total & 10 & 5 resolved, 3 residual, and 2 out of scope. \\
    \bottomrule
  \end{tabular}
  \caption{Gaps registered in the case study by severity and treatment.}
  \label{tab:atm-gaps}
\end{table}

In reconstruction, the migration plan was organized into 11 tasks. At inventory time, 9 were completed, the Docker parity task was in progress, and the final cutover stage remained pending, as summarized in Table~\ref{tab:atm-reconstruction}. This partial execution prevents strong conclusions about final migration success, but shows that the generated artifacts structured a Go implementation sequence with packages, SQLite persistence, entry points, parity tests, and technical handoff.

\begin{table}[t]
  \centering
  \footnotesize
  \begin{tabular}{p{6.3cm} p{2.3cm} p{3.4cm}}
    \toprule
    \textbf{Task group} & \textbf{Status} & \textbf{Produced evidence} \\
    \midrule
    Bootstrap, utilities, keyboard, repository, statement, account, logger, menu, and CLI entry & 9 completed & Go packages, SQLite, entry point, and domain modules. \\
    Parity with Docker and cross-execution & In progress & 53 Gherkin scenarios and Parallel Run strategy. \\
    Cutover, final validation, and Go release & Pending & Cutover plan and technical handoff. \\
    \bottomrule
  \end{tabular}
  \caption{Summary status of reconstruction tasks at inventory time.}
  \label{tab:atm-reconstruction}
\end{table}

\subsection{Case interpretation}
\label{subsec:atm-interpretation}

The case study offers three preliminary observations. First, Reversa was able to decompose a legacy system without tests into verifiable artifacts of domain, architecture, gaps, confidence, and parity. Second, confidence marking helped separate confirmed behavior from inference and absence of evidence. Third, the migration stage reused reverse documentation to produce a plan, architectural decisions, and executable scenarios.

These findings should be interpreted cautiously. The case has only one system, a simple domain, a single stakeholder, and an educational COBOL-to-Go migration. There is no controlled comparison against a single agent, a conventional documentation tool, or execution without specifications. Therefore, the case reinforces the plausibility of the framework and exemplifies its metrics, but does not replace the comparative studies proposed in Section~\ref{sec:evaluation-methodology}.

\section{Discussion}
\label{sec:discussion}

Reversa starts from a simple observation: coding agents can operate on real repositories, but their utility depends on the quality of the context they receive. In legacy systems, this context is rarely available as an explicit specification. The contribution of Reversa is to treat recovery of this context as a proper engineering stage, not as an incidental instruction before asking an agent to modify the system.

Traditional documentation and operational specification for agents have overlapping, but not identical, objectives. Documentation aimed at humans can be narrative, selective, or pedagogical. A coding agent needs more actionable contracts: behaviors to preserve, evidence that supports claims, inferred rules, gaps that block safe implementation, and tasks derived from each analyzed unit. Therefore, Reversa emphasizes traceability, confidence, and gaps. A fluent text may be comfortable for human reading and still be dangerous as input to automation if it presents fragile inferences as facts.

The multi-agent decomposition in Reversa does not claim that multiple agents are always superior to a single agent. The reason for separating Scout, Archaeologist, Detective, Architect, Writer, and Reviewer is to create checkpoints between different tasks: mapping the surface, extracting rules, synthesizing architecture, writing specifications, and reviewing confidence. This separation makes the process more auditable, because an incorrect claim can be traced to the stage in which it was introduced.

The COBOL-to-Go ATM study illustrates this difference. The pipeline did not only produce narrative documentation; it classified claims by confidence, registered gaps, organized human decisions, and derived parity scenarios. Although exploratory and partial, the case shows that the main output of Reversa is an operational basis for acting on the legacy system, not a static report separated from migration.

The paper should be read as the presentation of a framework, an evaluation protocol, and an exploratory case study, not as a conclusive demonstration of effectiveness. Artifact quality depends on agents, prompts, the chosen engine, the analyzed project, and the availability of evidence. Projects with highly dynamic code, rules hidden in databases, or production-dependent behavior may require additional instruments. Reversa does not remove these difficulties; it offers a way to make them visible, traceable, and treatable.

\section{Threats to Validity}
\label{sec:threats-validity}

This section summarizes threats to the validity of the study and delimits the scope of the conclusions. Because the paper presents a framework with a first exploratory instantiation, these threats are central to interpreting the results.

\textbf{Internal validity.} The ATM study was conducted without a controlled comparison against a single agent, a conventional documentation tool, or an execution without prior specifications. Therefore, it is not possible to isolate how much of the observed progress resulted from Reversa, stakeholder knowledge, domain simplicity, or human decisions made during migration. The internal confidence index was also computed from the classification produced by the pipeline itself, not by an independent audit.

\textbf{Construct validity.} Metrics such as confidence distribution, gaps, and completed tasks capture operational properties of the artifacts, but they do not directly measure factual accuracy, utility for agents, or final migration success. In the ATM case, expert precision and utility in downstream tasks were proposed as protocol metrics, but have not yet been measured in a controlled way.

\textbf{External validity.} The analyzed case involves an educational system in a simplified banking domain, with a single stakeholder, no production use, no preexisting tests, and a reduced scope of COBOL modules. Therefore, the results should not be generalized automatically to industrial legacy systems, regulated domains, larger teams, codebases with production-dependent behavior, or projects with complex external integrations.

\textbf{Conclusion validity.} The Go reconstruction was partial at inventory time: 9 of 11 tasks were completed, parity with Docker was in progress, and cutover remained pending. Consequently, the study supports only exploratory evidence of plausibility and process structuring, not conclusive evidence of final parity, cost reduction, improved agent performance, or superiority over alternatives.

\section{Conclusion}
\label{sec:conclusion}

Legacy systems concentrate operational knowledge that frequently does not exist as explicit specification. At the same time, coding agents depend on context, correctness criteria, and behavioral contracts to perform maintenance, migration, or evolution with lower risk. This combination creates a practical gap: before automating changes in legacy systems, it is necessary to recover and organize operational knowledge in a traceable and reviewable way.

This paper presented Reversa, a reverse documentation engineering framework for transforming legacy software into traceable operational specifications for AI agents. The proposal combines an installable CLI, support for multiple agent engines, specialized agentic teams, artifacts under \texttt{\_reversa\_sdd/}, SHA-256 manifest-based preservation, and an explicit model of confidence, inference, and gaps. We also reported an exploratory case study on migrating an ATM from COBOL to Go, in which the pipeline produced specifications, gaps, a confidence report, parity scenarios, and a reconstruction plan. Final parity and cutover were not completed in this version of the study. The central contribution is not to generate more documentation, but to reposition recovered documentation as an operational contract for agents that will act on the system.

The paper makes five contributions. First, we define reverse documentation engineering in the context of legacy systems and AI agents. Second, we describe Reversa as a multi-agent framework for converting existing code into operational specifications. Third, we present a confidence and gaps model to prevent fragile inferences from being treated as facts. Fourth, we detail an architecture that is portable across engines and preserves the legacy project. Fifth, we propose an evaluation protocol and instantiate it in an exploratory COBOL--Go case study.

The main limitation of this version is the absence of broad empirical validation. Therefore, Reversa should be interpreted as a framework with first exploratory evidence, not as conclusive evidence of superiority over alternative approaches. Next steps are to execute additional case studies in legacy projects, evaluate the artifacts with experts, compare the framework against single-agent and conventional-documentation baselines, measure utility in downstream tasks, and refine the metrics for coverage, traceability, confidence, and cost.

As future work, we plan to extend the initial study to legacy projects across different domains, languages, and degrees of prior documentation. This extension should include auditable artifact packages, independent artifact review, controlled comparison with baselines, downstream tasks executed by coding agents, and analysis of the cost of maintaining specifications over new changes. The goal is to transform the proposal presented here into systematic evidence about the role of reverse documentation engineering in AI-agent-assisted maintenance, migration, and evolution.

\section*{Generative AI Use Statement}

The author conducted the research and wrote the manuscript. During the preparation of this study, however, the author used Grammarly tools to improve textual agreement and OpenAI/Codex to support text structuring and translation into English. After using these tools/services, the author reviewed and edited the content as needed and takes full responsibility for the content of the publication.

\bibliographystyle{plain}
\bibliography{refs}

\end{document}